\documentclass[aip,amsmath,amssymb,reprint]{revtex4-1}
\usepackage[english]{babel}
\usepackage{graphicx}
\usepackage{dcolumn}
\usepackage{bm}
\usepackage[utf8]{inputenc}
\usepackage[T1]{fontenc}
\usepackage{mathptmx}
\usepackage{etoolbox}
\usepackage{appendix}
\usepackage{xcolor}

\draft
\begin{document}
\preprint{AIP/123-QED}

\title[Reducing TLS loss in tantalum resonators using titanium sacrificial layers]{Reducing TLS loss in tantalum CPW resonators using titanium sacrificial layers}

\author{Zachary Degnan}
 \email{z.degnan@uq.edu.au}
\author{Chun-Ching Chiu}
\affiliation{School of Mathematics and Physics, The University of Queensland, Brisbane, QLD 4072, Australia}

\author{Yi-Hsun Chen}%
\affiliation{School of Mathematics and Physics, The University of Queensland, Brisbane, QLD 4072, Australia}

\author{David Sommers}%
\affiliation{School of Mathematics and Physics, The University of Queensland, Brisbane, QLD 4072, Australia}

\author{Leonid Abdurakhimov}
\affiliation{IQM Finland, Keilaranta 19, Espoo 02159, Finland}

\author{Lihuang Zhu}
\affiliation{IQM Finland, Keilaranta 19, Espoo 02159, Finland}

\author{Arkady Fedorov}
 \email{a.fedorov@uq.edu.au}
\author{Peter Jacobson}
 \email{p.jacobson@uq.edu.au}
\affiliation{School of Mathematics and Physics, The University of Queensland, Brisbane, QLD 4072, Australia}

\date{\today}

\begin{abstract}
We demonstrate a substantial reduction in two-level system loss in tantalum coplanar waveguide resonators fabricated on high-resistivity silicon substrates through the use of an ultrathin titanium sacrificial layer. 
A 2~Å titanium film, deposited atop pre-sputtered $\alpha$-tantalum, acts as a solid-state oxygen getter that chemically modifies the native Ta oxide at the metal–air interface. 
After device fabrication, the titanium layer is removed using buffered oxide etchant, leaving behind a chemically reduced Ta oxide surface. 
Subsequent high-vacuum annealing further suppresses two-level system loss. Resonators treated with this process exhibit internal quality factors $Q_i$ exceeding an average of 1.5 million in the single-photon regime across ten devices—over three times higher than otherwise identical devices lacking the titanium layer. 
These results highlight the critical role of interfacial oxide chemistry in superconducting loss and reinforce atomic-scale surface engineering as an effective approach to improving coherence in tantalum-based quantum circuits. 
The method is compatible with existing fabrication workflows applicable to tantalum films, offering a practical route to further extending $T_1$ lifetimes in superconducting qubits.
\end{abstract}

\maketitle

Reducing loss and decoherence in superconducting circuits through materials engineering is a direct and scalable approach to enhancing qubit coherence, compared to alternative circuit-level approaches such as participation ratio optimisation.~\cite{wang_surface_2015, calusine_analysis_2018, martinis_surface_2022, guo_near-field_2023}
By targeting the microscopic origins of decoherence~\cite{müller_understanding_2019, guo_near-field_2023}—particularly dielectric loss from two-level systems (TLS) at device interfaces~\cite{woods_determining_2019, crowley_disentangling_2023, mclellan_chemical_2023}—this strategy enables intrinsic improvements to $T_1$ and $T_2$ without requiring changes to qubit architecture. 
Crucially, materials-based methods have the advantage of being readily compatible with existing device layouts and fabrication workflows, making them broadly accessible. 

The adoption of tantalum (Ta) as a superconducting material has led to major advances in the coherence times of superconducting quantum processors and internal quality factors of superconducting resonators,~\cite{ganjam_surpassing_2024, wang_practical_2022, place_new_2021, shi_tantalum_2022, goronzy_comparison_2025, pritchard_suppressed_2025} including recent demonstrations of 2D transmon qubits with energy relaxation times ($T_1$) exceeding 1.6 milliseconds.~\cite{bland_2d_2025}
This improvement is widely attributed to the chemically stable and high-quality native oxide formed on Ta, which plays a key role in suppressing dielectric loss from TLS.~\cite{crowley_disentangling_2023, place_new_2021} 
This confers a distinct advantage over other superconductors such as niobium (Nb), whose native oxide is comparatively lossy.~\cite{verjauw_investigation_2021, bal_systematic_2024}
However, despite the reduced intrinsic losses of Ta oxides, Ta-on-silicon superconducting resonators and qubits remain limited by TLS losses originating from surface oxides---particularly at the metal–air (MA) interface---underscoring the need for a more precise understanding and control of interfacial oxide chemistry and structure.~\cite{bland_2d_2025, lozano_low-loss_2024}
Amorphous Ta pentoxide (a-Ta\textsubscript{2}O\textsubscript{5})---the primary native oxide phase---grows spontaneously upon air exposure, owing to the highly negative formation energy of Ta\textsubscript{2}O\textsubscript{5}.~\cite{jacob_update_2009}
Although thinner than native oxides on other superconductors, such as NbO\textsubscript{x}, the Ta oxide still contributes significantly to decoherence through TLS dipoles associated with its amorphous structure.~\cite{crowley_disentangling_2023}

\begin{figure*}
    \includegraphics[width=\linewidth]{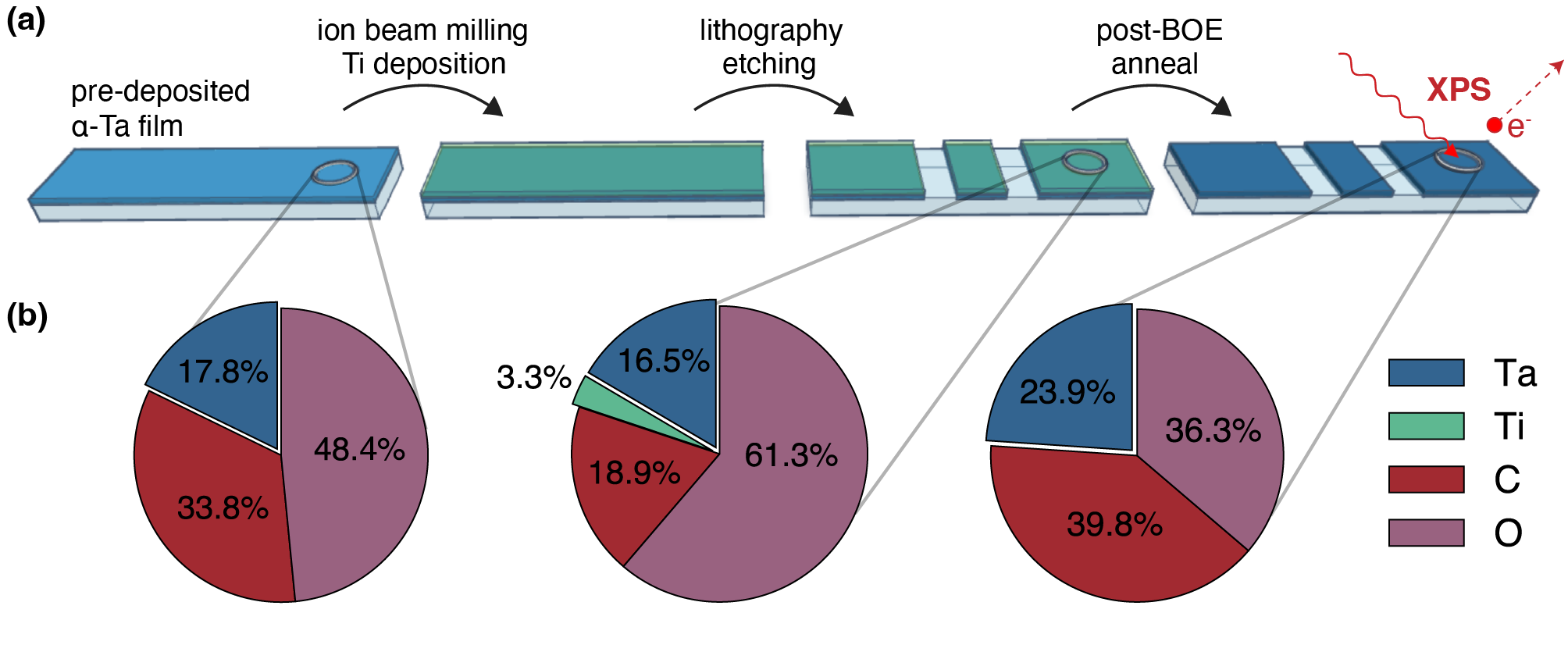}
    \caption{\label{fig:process}\textbf{(a)} Schematic of CPW fabrication with a sacrificial Ti layer. Native Ta oxide was removed by Ar$^+$ ion beam milling, followed by \textit{in situ} deposition of a 2~Å Ti layer. Circuit patterning was performed using positive-tone optical lithography and CF$_4$-based reactive ion etching. Post-fabrication treatments consisted of buffered oxide etching (BOE) and annealing. \textbf{(b)} Surface atomic concentrations from XPS survey scans at stages marked in (a). The Ti layer is retained during lithography and etching, then removed by long-BOE prior to annealing.}
\end{figure*}

While oxides can be readily removed through techniques such as chemical or physical etching, their spontaneous regrowth is unavoidable during fabrication and device packaging.
This has motivated studies on capping and passivation layers---such as gold, gold-palladium, magnesium oxide and organic self-assembled monolayers---deposited atop superconducting films to prevent or minimise oxide regrowth.~\cite{chang_eliminating_2025, alghadeer_surface_2023, zhou_ultrathin_2024}
These methods are generally successful in reducing oxide regrowth and improving surface metallicity, however the resulting reductions in TLS loss are modest at best, and in some cases increased due to the added interface complexity on the measured devices.

In parallel to capping- or passivation-layers, sacrificial layers offer an approach to modify the underlying film surface without remaining in-place during device operation, typically being removed by selective etching at the end of device fabrication.
Materials such as yttrium (Y)~\cite{wang_reducing_2017} and titanium (Ti)~\cite{joiner_cleaning_2014} have been employed as sacrificial layers---up to 10~nm in thickness---in high-mobility graphene devices to improve cleanliness, simplify large-scale fabrication, and reduce device-to-device variability. 
Deposited prior to critical fabrication steps and subsequently removed, the sacrificial layer serves to getter oxides and contaminants, and protect the active surface from residue accumulation.

Here, we demonstrate a surface engineering approach that enhances the quality factor of superconducting coplanar waveguide (CPW) resonators by modifying the Ta oxide through the application of an ultrathin Ti sacrificial layer, significantly reducing TLS loss at the MA interface and reducing device variability.
Fabricating devices with such a sacrificial Ti layer, along with post-fabrication buffered oxide etching (BOE) and high temperature annealing, we observe up to a four-fold increase in the internal Q-factor ($Q_i$) of Ta-on-Si CPW resonators compared to otherwise identical devices without the Ti layer. 
These resonators reach values above 2 million at single photon occupancy, and above 10 million at higher photon numbers. 

Unlike traditional capping layers, which remain on the device and can complicate fabrication or introduce additional interfaces,~\cite{bal_systematic_2024, chang_eliminating_2025, zhou_ultrathin_2024} our method uses an ultrathin Ti film as a temporary, removable, and reactive surface modifier. 
Deposited on a Ta film, the Ti layer acts as a solid-state oxygen getter, preferentially reacting with ambient or residual oxygen to modify the formation dynamics of the native Ta oxide. 
This alters the interfacial chemistry, leading to a thinner, less defective Ta oxide with reduced TLS density, and therefore a lower MA loss tangent $\tan\delta^\text{ TLS}_\text{ MA}$. 
Furthermore, the Ti layer has a dual-use of acting as a kinetic barrier to (re)oxidation, protecting the underlying Ta during lithographic fabrication steps.
From the phase diagram of the Ta-Ti system, it can be seen that alloys are unable to form below 882~$^\circ$C,~\cite{murray_material_2021} preventing detrimental inter-diffusion of the layers during annealing steps during fabrication.
After lithography, the Ti is subsequently removed using buffered oxide etchant (BOE), eliminating it from the final device while retaining the favourable oxide modification. 
The process offers a clean and scalable route to improving dielectric interfaces and is highly compatible with standard thin-film deposition and microfabrication techniques, enabling rapid adoption by groups seeking to reduce surface losses in superconducting quantum circuits.

\begin{figure}
    \centering
    \includegraphics[width=1\linewidth]{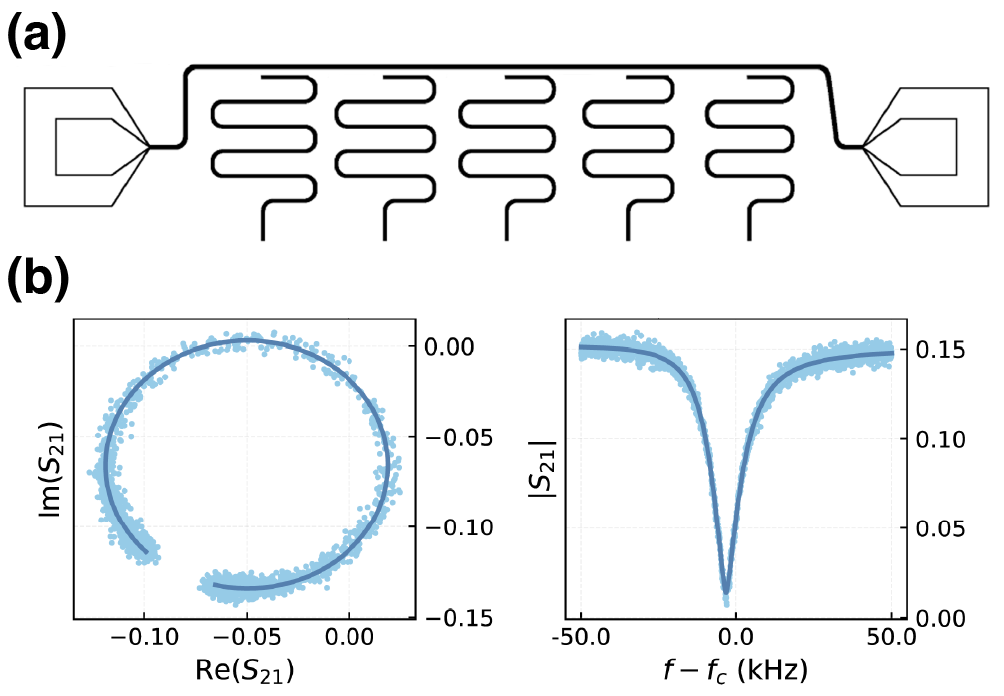}
    \caption{\textbf{(a)} Schematic of the fabricated circuit with five $\lambda/4$ resonators side-coupled to a transmission line. The resonator CPW has a 5~µm gap and 9~µm width. \textbf{(b)} Complex $S_{21}$ data (points) and circle fit (line) over a 100~kHz span around the centre frequency $f_c$. The scattering data traces a circle in the complex plane (left) and produces a Lorentzian dip in amplitude (right)~\cite{rieger_fano_2023}.}
    \label{fig:wiring_circuit}
\end{figure}

\begin{figure*}
    \includegraphics[width=\linewidth]{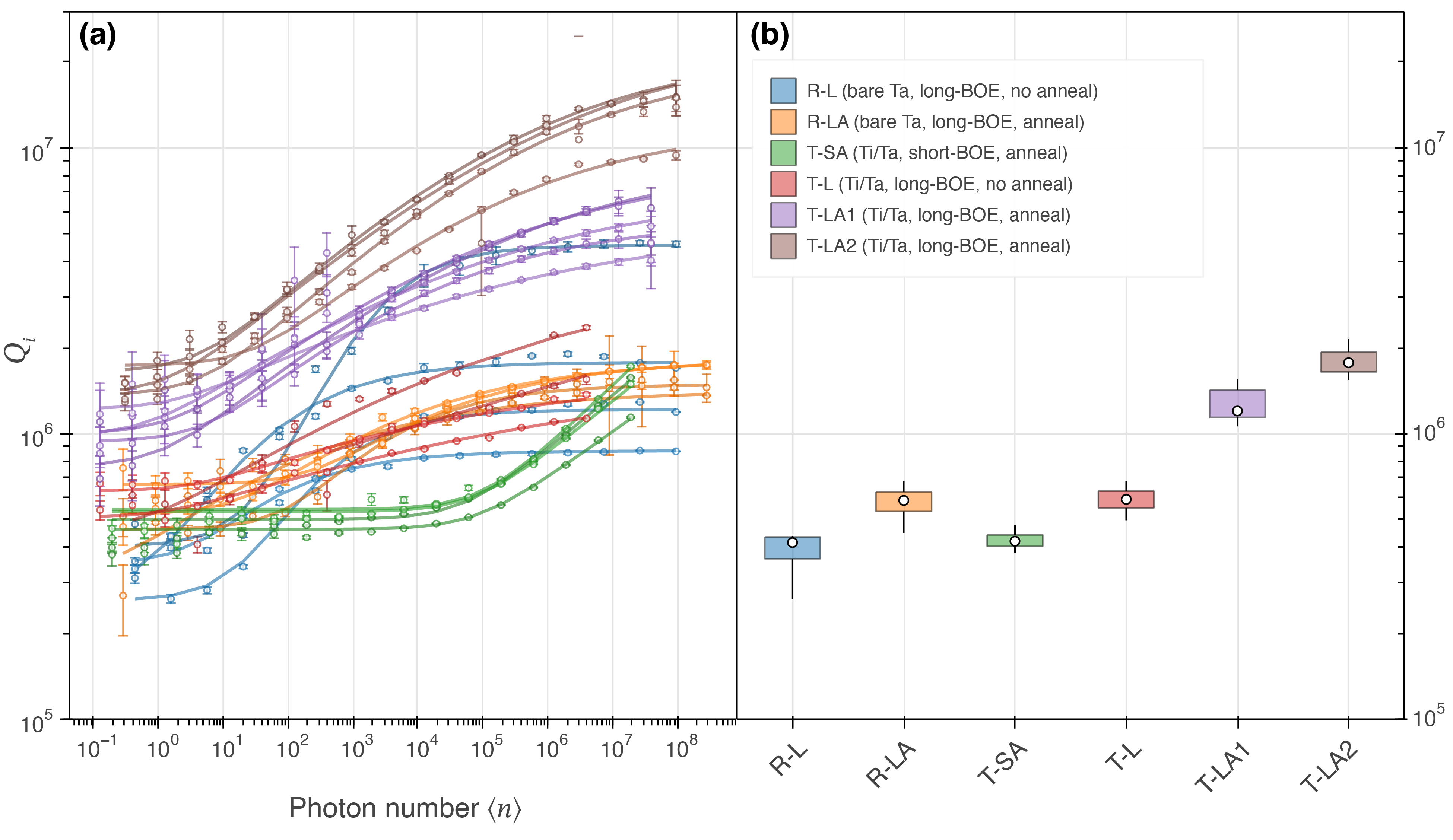}
    \caption{\label{fig:Qi_comparison} Internal quality factor measurements for bare Ta reference samples, and variants of Ti/Ta devices. \textbf{(a)} Measured $Q_i$ data (points) as a function of mean intracavity photon number $\langle n \rangle$ with a saturable TLS model fit according to Eq.~\ref{eq:TLS_model} (solid lines), yielding $F\delta_\text{TLS}$ values given in Table~\ref{tab:QR_results}. \textbf{(b)} Box plot showing the distribution of $Q_i$ values at single photon power ($\langle n \rangle\approx1$) across resonators on each sample. The box spans the interquartile range (IQR) from the 25th (Q1) to the 75th (Q3) percentile, with a white dot at the median value. Whiskers extend to the most extreme data points within 1.5$\times$IQR from the box.}
\end{figure*}

To assess the impact of the Ti sacrificial layer and to separate its contribution from other fabrication-dependent effects, we fabricate CPW resonators from 200~nm sputtered $\alpha$-Ta films on high-resistivity Si. The devices were designed and simulated using the open-source SQDMetal software stack~\cite{sommers_open-source_2025}. A subset of chips are produced on films incorporating a 2~Å Ti sacrificial layer, deposited prior to any patterning (see Appendix for details). Following optical lithography and CF$_4$ reactive ion etching, all chips undergo resist stripping and an O$_2$-plasma clean, which leaves both Ta and Si surfaces oxidised. After their final BOE treatment (described below), samples are wire-bonded and packaged within hours to minimise regrowth of surface oxides before cryostat loading.

To systematically explore how surface preparation influences resonator performance, we fabricate devices spanning a controlled range of post-processing parameters. Each chip receives some combination of: (i) BOE exposure—either a short 1~min etch (ID: S) or a long 20~min etch (ID: L); (ii) high-temperature annealing at $700^\circ\mathrm{C}$ for 10~h (ID: A) or no anneal; and (iii) inclusion (ID: T) or omission (ID: R) of the 2~Å Ti sacrificial layer. These IDs correspond to the device labels used in Table~\ref{tab:QR_results} and Fig.~\ref{fig:Qi_comparison}. By independently varying BOE duration, annealing, and Ti-layer inclusion, we isolate the effect of each treatment on the resulting resonator quality.

\begin{table*}
    \small
    \centering
    \setlength{\tabcolsep}{6pt}
    \vspace{0.5em}
    \begin{tabular}{|l||lll|cccc|}
        \hline
        \multicolumn{1}{|c||}{~} & 
        \multicolumn{3}{c|}{Fabrication Treatments} &
        \multicolumn{4}{c|}{Loss Values} \\
        \hline\hline
        Sample ID & Ti layer & Post-BOE & Anneal &
        \boldmath$F\delta_{\text{TLS}}\,(\times10^{-6})$  &
        \boldmath$Q_{i,\text{max}}^{\text{LP}}\,(\times10^6)$ &
        \boldmath$\bar{Q}_i^{\text{LP}}\,(\times10^6)$ &
        \boldmath$\bar{Q}_i^{\text{HP}}\,(\times10^6)$ \\
        \hline
        \texttt{R-L}   & None & 20~min & None &
            $2.86 \pm 0.41$ & $0.44$ & $0.42 \pm 0.03$ & $2.29 \pm 0.65$ \\
        \texttt{R-LA}  & None & 20~min & 700$^\circ$C, 10 hr &
            $1.11 \pm 0.19$ & $0.66$ & $0.57 \pm 0.09$ & $1.57 \pm 0.04$ \\
        \texttt{T-SA}  & 2~Å & 1~min & 700$^\circ$C, 10 hr &
            $2.26 \pm 0.06$ & $0.48$ & $0.43 \pm 0.07$ & $1.48 \pm 0.29$ \\
        \texttt{T-L}   & 2~Å & 20~min & None &
            $1.17 \pm 0.20$ & $0.68$ & $0.58 \pm 0.05$ & $1.68 \pm 0.35$ \\
        \texttt{T-LA1} & 2~Å & 20~min & 700$^\circ$C, 10 hr &
            $0.89 \pm 0.05$ & $1.55$ & $1.27 \pm 0.09$ & $5.38 \pm 0.46$ \\
        \texttt{T-LA2} & 2~Å & 20~min & 700$^\circ$C, 10 hr &
            $0.58 \pm 0.03$ & $2.14$ & $1.81 \pm 0.13$ & $13.36 \pm 1.34$ \\
        \hline
    \end{tabular}
    \caption{$Q_i$ measurement results for resonator devices and filling-factor-adjusted TLS loss tangents $F\tan\delta_\text{TLS}$ extracted from power-dependent $Q_i$ data (Fig.~\ref{fig:Qi_comparison}). Sample IDs follow the format X-BA, where X denotes the material type (R = bare Ta, T = Ti-treated), B indicates the BOE treatment (L = long, S = short), and A marks whether the sample was annealed (A = annealed, omitted if not). Values are shown as the mean taken across all resonators on the sample, except $Q_{i,\text{max}}^{\text{LP}}$ which is the maximum single photon $Q_i$ on each sample. The standard deviation on the average is also shown. Superscripts HP and LP refer to low- and high-power respectively, where $\langle n \rangle^\text{LP}\approx1$, and $\langle n \rangle^\text{HP}\geq10^6$. We note that losses in some samples losses had not yet saturated (i.e., reached a steady state with respect to photon number) at high power.}
    \label{tab:QR_results}
\end{table*}

Each device contains five $\lambda/4$ CPW resonators capacitively coupled to a common feed-line, as shown in Fig.~\ref{fig:wiring_circuit}(a). Complex $S_{21}(f)$ transmission data was recorded in a 0.5~MHz span around the fundamental frequency $f_0$ of each resonator, designed to be between 5.5~GHz and 7.0~GHz for all samples.
For power-dependent $Q_i$ measurements, as shown in Fig.~\ref{fig:Qi_comparison}, the applied power $P_\text{app}$ at the resonator input was varied from $-72$~dBm to $-162$~dBm, corresponding to average intracavity photon numbers $\langle n \rangle$ from $10^{-1}$ to $10^8$.
Complex $S_{21}$ resonance spectra---such as that shown in Fig.~\ref{fig:wiring_circuit}(b)---were fit using a diameter-corrected circle-fit method developed by Probst \textit{et al.}, allowing for the extraction of $Q_i$, and the coupling quality factor $Q_c$.~\cite{probst_efficient_2015}

The magnitude of TLS loss $\delta_\text{TLS}$ was determined using a model accounting for losses due to TLSs ($\delta_\text{TLS}(\langle n \rangle, \ T)$), quasiparticles ($\delta_\text{QP}(T)$), and temperature- and power-independent high power losses ($\delta_\text{HP}$):~\cite{crowley_disentangling_2023}
\begin{equation}
    \frac{1}{Q_i(\langle n \rangle, \ T)}=\delta_\text{TLS}(\langle n \rangle, \ T) + \delta_\text{QP}(T)+\delta_\text{other}.
    \label{eq:total_losses}
\end{equation}
The contributions from TLS losses are parametrised by
\begin{equation}
    \delta_\text{TLS}(\langle n \rangle, \ T)=F\delta^0_\text{TLS}\frac{\tanh\left(\frac{\hslash\omega_0}{2k_B T}\right)}{\left(1+\frac{\langle n \rangle}{n_c}\right)^\beta},
    \label{eq:TLS_model}
\end{equation}
where $F$ is the geometry-dependent filling factor, $\delta^0_\text{TLS}$ is the linear TLS absorption, $\omega_0=2\pi f_0$ is the fundamental angular frequency of the resonator, and $n_c$ is the critical photon number above which TLS saturate, and $\beta$ is an exponent varying between $\beta=1/2$ for non-interacting TLS, and $0.5<\beta\leq0$ for interacting TLS.~\cite{gao_physics_2008, de_graaf_suppression_2018} 
All power sweeps were taken at a constant temperature of $T\approx25.7$~mK.

\begin{figure}
    \centering
    \includegraphics[width=1\linewidth]{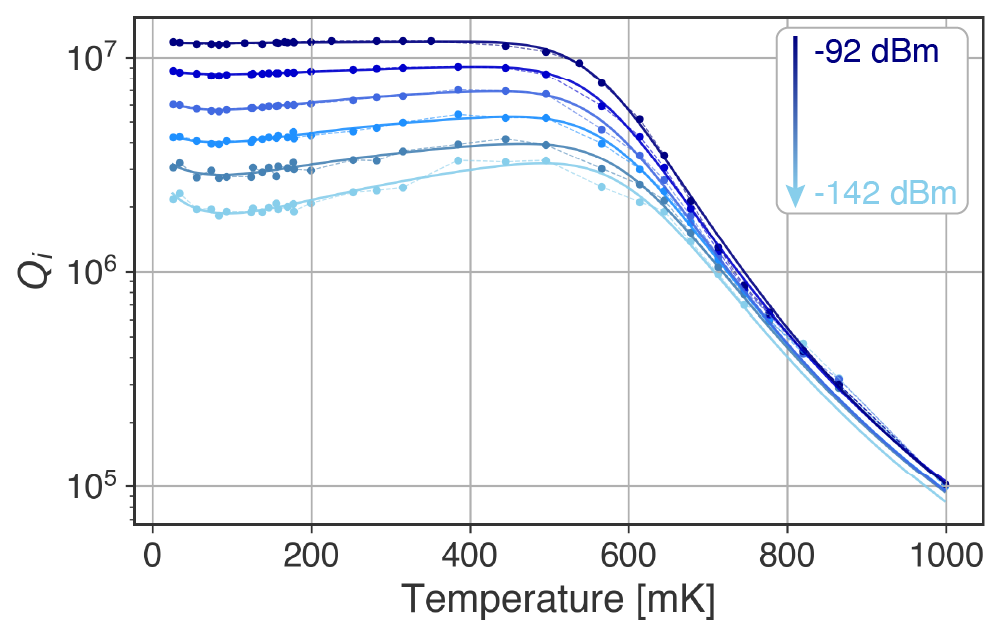}
    \caption{Temperature dependence of $Q_i$ for a single resonator at $f_0=6.34$~GHz on a Ti-treated and annealed sample (\texttt{T-LA2}). Experimental data is shown as points, with a fitted temperature- and power-dependent loss model (Eq.~\ref{eq:total_losses}) displayed as a solid line. The measurement was repeated at six evenly-spaced power levels between $P_\text{app}=-92$~dBm (dark blue) and $P_\text{app}=-142$~dBm (light blue), corresponding to intracavity photon numbers of $\langle n \rangle \approx 10^6$ and $\langle n \rangle \approx 10^1$ respectively.}
    \label{fig:temp}
\end{figure}

Temperature-dependent measurements of $Q_i$ are shown in Fig.~\ref{fig:temp}. The temperature dependence of TLS and thermal quasiparticle losses was quantified by extracting $Q_i$ over a range from $T = 25$~mK to $T = 998$~mK, at six photon occupancies spanning approximately $\langle n \rangle \approx 10^1$ to $\langle n \rangle \approx 10^6$.
Temperature-dependent losses due to thermally excited quasiparticles are modelled as 
\begin{equation}
    \delta_\text{QP}(T)=\delta_\text{QP}^0\frac{\sinh\left(\frac{\hslash\omega_0}{2k_B T}\right)K_0\left(\frac{\hslash\omega_0}{2k_B T}\right)}{e^{\Delta_0/k_B T}},
    \label{eq:equilibrium_QPs}
\end{equation}
where $\delta_\text{QP}^0$ is the linear absorption from thermal quasiparticles, $\Delta_0$ is the superconducting gap ($\Delta_0=1.764 k_B T_C$, and $T_C=4.4$~K is the superconducting critical temperature of the film), and $K_0$ is the zeroth order modified Bessel function of the second kind.~\cite{gao_physics_2008, crowley_disentangling_2023}

Figure~\ref{fig:Qi_comparison} shows the power-dependent and single-photon $Q_i$ for devices fabricated with variations in post-lithography treatment (short/long BOE), annealing, and the inclusion of a 2~Å Ti layer.
We use sample~\texttt{R-L} as a benchmark, fabricated using standard methods for high-$Q$ Ta-on-Si CPW resonators (bare Ta, long BOE, no anneal)~\cite{lozano_manufacturing_2022, crowley_disentangling_2023}, which exhibits a mean $Q_i \approx 4.2 \times 10^5$ at single-photon power.
This device serves as a baseline for comparison, with its internal quality factor reflecting the intrinsic properties of our Ta-on-Si films.
While not the focus of this study, our baseline $Q_i$ is slightly lower than reported state-of-the-art resonators fabricated by similar methods (including post-lithography BOE/HF treatment) on optimised $\alpha$-Ta films ($\sim$3 to 4 million)~\cite{lozano_low-loss_2024, marcaud_low-loss_2025}, and should be considered when evaluating the impact of the Ti sacrificial layer.
We first examine the impact of annealing on a bare Ta sample after a 20~min BOE treatment, observing a substantial increase in quality factor to $Q_i \approx 5.7 \times 10^5$, corresponding to a $\sim61\%$ reduction in TLS losses. 
This improvement is attributed to a favourable rearrangement of the local coordination environment at the MA interface, where annealing modifies both the Ta oxide and the underlying metal layer to suppress TLS formation.~\cite{crowley_disentangling_2023} 
The crystallisation of Ta\textsubscript{2}O\textsubscript{5} films has been reported to onset between 700$^\circ$C and 800$^\circ$C, suggesting that regions of our oxide may begin to crystallise during the 10~hr anneal at 700$^\circ$C,~\cite{ren_annealing_2021, xu_high_2008} however we were unable to confirm the existence of crystalline regions with grazing incidence X-ray scattering experiments. 
Similar restructuring and crystallisation of surface oxides have been observed in niobium devices, where they correlate with reduced losses.~\cite{kalboussi_reducing_2024}
Annealing consistently improves performance in both bare Ta and Ti-treated devices, and ongoing efforts aim to optimise the thermal budget of this step by lowering its duration.
Devices \texttt{R-LA} (bare Ta, long-BOE, anneal), \texttt{T-SA} (2~Å Ti, short-BOE, anneal), and \texttt{T-L} (2~Å Ti, long-BOE, no anneal) exhibit similar low power losses to the baseline sample. 
Consistent with previous studies, we find that the post-lithography BOE step substantially reduces saturable losses, primarily by removing lossy  SiO\textsubscript{x} at the SA interface and partially thinning the native Ta oxide at the MA interface.
Lozano \emph{et al.} recently reported a similar effect in Ta-on-Si devices, showing that HF etching significantly suppresses loss contributions from SiO\textsubscript{x} at the SA interface, resulting in a 2--4$\times$ reduction in total TLS losses.~\cite{lozano_low-loss_2024}
In our devices, comparing samples \texttt{T-SA} (short 1~min BOE) and \texttt{T-LA} (long 20~min BOE), we likewise observe a 2--4$\times$ reduction in $F\delta_\mathrm{TLS}$.
We also observe that the device exposed to only 1~min BOE (\texttt{T-SA}) has a much higher critical photon number $n_c$ (see Eq.~\ref{eq:TLS_model}), suggesting a larger number of depolarising TLS.
As confirmed by XPS measurements and shown in Fig.~\ref{fig:process}, the 2~Å Ti sacrificial layer is fully removed after the long BOE treatment.
By contrast, XPS on a sample equivalent to \texttt{T-SA} indicates that the Ti layer remains intact after the short BOE step (see Appendix).
While BOE effectively reduces losses associated with TLS hosted at the SA interface, its influence on the MA interface is limited.~\cite{lozano_low-loss_2024}
To directly address losses at the MA interface, we introduce a Ti sacrificial layer as a complementary strategy. 
Comparing the bare-Ta reference sample \texttt{R-LA} with the Ti-treated samples \texttt{T-LA1} and \texttt{T-LA2}, we observe an additional enhancement in the single-photon $Q_i$ by a factor of $\sim$2--4, which we attribute to loss reduction enabled by the sacrificial layer. 
At high power, the $Q_i$ of the Ti-treated devices further improves by a factor of $\sim$3--9 relative to the reference. 
The magnitude of these improvements is comparable to those achieved through long-BOE treatment, which is a well-established key step in the fabrication of ultra-high-$Q$ Ta resonators. 
Taken together, these results demonstrate that the Ti sacrificial layer substantially enhances an already optimised Ta fabrication process, providing a clear pathway to longer coherence times in state-of-the-art superconducting qubits.

Devices with residual Ti (sample \texttt{T-SA}) exhibited reduced $Q_i$, indicative of additional loss channels likely associated with an amorphous TiO\textsubscript{x} layer at the MA interface. 
This highlights the critical role of precise chemical control and selectivity in achieving oxide suppression, and clarifies that the Ti layer does not function as a capping layer.

Despite concerns from prior work that ion milling can degrade $Q_i$ by restructuring oxides and introducing oxygen defects,\cite{van_damme_argon-milling-induced_2023, arabi_magnetic_2025} we observe no performance loss following the argon ion milling step used to remove native Ta oxides prior to Ti deposition.
We attribute this resilience to the oxygen-scavenging action of the Ti layer, which likely removes embedded defects, and to the beneficial effects of subsequent annealing, which may passivate surfaces or recrystallise near-surface regions.\cite{kalboussi_reducing_2024}
Notably, even the un-annealed Ti/Ta sample (argon-milled) outperforms the bare Ta control (not milled), suggesting that the Ti layer alone substantially mitigates milling-induced damage.

Temperature-dependent measurements, as shown in Fig.~\ref{fig:temp}, demonstrate that quasiparticle losses become appreciable only at elevated temperatures ($T \gtrsim 400$~mK, where the thermal energy $k_BT$ is an appreciable fraction of the intracavity photon energy $\hbar\omega_0$), confirming that losses at qubit operating temperatures are dominated by dielectric TLSs. 
The extracted TLS loss tangents are near the lower bound expected for this material system and device geometry. 
Compared to state-of-the-art devices reported in the literature, such as the compilation by McRae \emph{et al.}, our resonators approach the performance limit for sub-10~µm features on silicon substrates.~\cite{mcrae_materials_2020}

The use of a Ti sacrificial layer represents a novel approach to improving the internal quality factor $Q_i$ of superconducting Ta-on-Si CPW resonators by addressing losses at the metal-air (MA) interface.
Unlike conventional capping layers that are co-deposited with the bulk film and retained throughout processing, this method is applied \emph{post-deposition} and fully removed prior to measurement.
This decoupling of growth and oxide engineering enables greater flexibility and compatibility with standard fabrication workflows.
As Ti is readily available in most electron beam evaporation systems, the technique is also broadly accessible.
While demonstrated here on CPW resonators, the method is directly transferable to base-layer fabrication in planar qubit architectures such as the transmon.

In conclusion, we have demonstrated a Ti sacrificial layer approach for reducing losses in Ta-based superconducting circuits. 
Whereas previous improvements to the quality factor of Ta resonators have largely targeted the SM and SA interfaces, our method directly mitigates losses at the MA interface. 
We fabricated and characterised a broad set of superconducting CPW resonators, enabling direct comparison of TLS losses between otherwise identical devices. 
XPS measurements were employed to track chemical changes at the MA interface during fabrication and across device variants. 
Devices incorporating a 2~Å sacrificial Ti layer consistently exhibited single-photon $Q_i$ above one million, representing a 2--4$\times$ enhancement compared to otherwise identical devices without the Ti layer. 
Temperature-dependent $Q_i$ measurements confirm that losses in the operating regime are dominated by depolarising TLS. 
To further improve process robustness and accessibility, future work should explore the use of thicker Ti layers, which may simplify process control and increase fabrication tolerance. 
The impact of annealing steps with reduced thermal budget, and their compatibility with higher-throughput fabrication, should also be investigated.

\section{Author contributions}
\textbf{Zachary Degnan}: Data curation (lead); Measurement (lead); Fabrication (lead); Simulation (equal); Conceptualisation (equal); Formal analysis (lead); XPS (lead); GIWAXS (lead); Investigation (equal); Validation (lead); Visualisation (lead); Writing – original draft (lead); Writing – review \& editing (lead). \textbf{Chun-Ching Chiu, Yi-Hsun Chen}: Measurement (supporting); Fabrication (supporting); XPS (supporting). \textbf{David Sommers}: Simulation (equal); Measurement (supporting). \textbf{Leonid Abdurakhimov, Lihuang Zhu}: Fabrication (supporting); Investigation (supporting); Writing – review \& editing (supporting). \textbf{Arkady Fedorov}: Investigation (equal); Funding acquisition (equal); Conceptualisation (equal). \textbf{Peter Jacobson}: Investigation (equal); Funding acquisition (equal); Conceptualisation (equal); Writing – original draft (supporting); Writing – review \& editing (supporting).

\section{Acknowledgements}
The authors acknowledge that UQ operates on the land of the Jagera and Turrbal peoples. The authors pay respects to their ancestors and descendants who continue to uphold connection to country. GIWAXS measurements were undertaken on the NCD-SWEET beamline at the ALBA Synchrotron (Barcelona, Spain). The authors acknowledge the assistance of Dr. Julian Steele and Dr. Eduardo Solano with GIWAXS measurements. We acknowledge travel funding provided by the International Synchrotron Access Program (ISAP) managed by the Australian Synchrotron, part of ANSTO, and funded by the Australian Government. The authors also acknowledge the facilities, and the scientific and technical assistance from staff at the Centre for Microscopy and Microanalysis (CMM) at the University of Queensland. This work used the Queensland node of the NCRIS-enabled Australian National Fabrication Facility (ANFF). This work was partially supported by the Australian Research Council under the grant LP210200636.

\bibliography{references}

@article{alghadeer_surface_2023,
  title = {Surface {{Passivation}} of {{Niobium Superconducting Quantum Circuits Using Self-Assembled Monolayers}}},
  author = {Alghadeer, Mohammed and Banerjee, Archan and Hajr, Ahmed and Hussein, Hussein and Fariborzi, Hossein and Rao, Saleem Ghaffar},
  year = 2023,
  month = jan,
  journal = {ACS Applied Materials \& Interfaces},
  volume = {15},
  number = {1},
  pages = {2319--2328},
  publisher = {American Chemical Society (ACS)},
  issn = {1944-8244, 1944-8252},
  doi = {10.1021/acsami.2c15667},
  urldate = {2025-07-24},
  langid = {english}
}

@article{arabi_magnetic_2025,
  title = {Magnetic Bound States Embedded in Tantalum Superconducting Thin Films},
  author = {Arabi, Soroush and Li, Qili and Dhundhwal, Ritika and Fuchs, Dirk and Reisinger, Thomas and Pop, Ioan M. and Wulfhekel, Wulf},
  year = 2025,
  month = mar,
  journal = {Applied Physics Letters},
  volume = {126},
  number = {11},
  pages = {114001},
  issn = {0003-6951, 1077-3118},
  doi = {10.1063/5.0251996},
  urldate = {2025-07-28},
  langid = {english}
}

@article{bal_systematic_2024,
  title = {Systematic Improvements in Transmon Qubit Coherence Enabled by Niobium Surface Encapsulation},
  author = {Bal, Mustafa and Murthy, Akshay A. and Zhu, Shaojiang and Crisa, Francesco and You, Xinyuan and Huang, Ziwen and Roy, Tanay and Lee, Jaeyel and Zanten, David Van and Pilipenko, Roman and Nekrashevich, Ivan and Lunin, Andrei and Bafia, Daniel and Krasnikova, Yulia and Kopas, Cameron J. and Lachman, Ella O. and Miller, Duncan and Mutus, Josh Y. and Reagor, Matthew J. and Cansizoglu, Hilal and Marshall, Jayss and Pappas, David P. and Vu, Kim and Yadavalli, Kameshwar and Oh, Jin-Su and Zhou, Lin and Kramer, Matthew J. and Lecocq, Florent and Goronzy, Dominic P. and {Torres-Castanedo}, Carlos G. and Pritchard, P. Graham and Dravid, Vinayak P. and Rondinelli, James M. and Bedzyk, Michael J. and Hersam, Mark C. and Zasadzinski, John and Koch, Jens and Sauls, James A. and Romanenko, Alexander and Grassellino, Anna},
  year = 2024,
  month = apr,
  journal = {npj Quantum Information},
  volume = {10},
  number = {1},
  pages = {43},
  issn = {2056-6387},
  doi = {10.1038/s41534-024-00840-x},
  urldate = {2024-12-06},
  langid = {english}
}

@misc{bland_2d_2025,
  title = {{{2D}} Transmons with Lifetimes and Coherence Times Exceeding 1 Millisecond},
  author = {Bland, Matthew P. and Bahrami, Faranak and Martinez, Jeronimo G. C. and Prestegaard, Paal H. and Smitham, Basil M. and Joshi, Atharv and Hedrick, Elizabeth and {Pakpour-Tabrizi}, Alex and Kumar, Shashwat and Jindal, Apoorv and Chang, Ray D. and Yang, Ambrose and Cheng, Guangming and Yao, Nan and Cava, Robert J. and {de Leon}, Nathalie P. and Houck, Andrew A.},
  year = 2025,
  month = mar,
  number = {arXiv:2503.14798},
  eprint = {2503.14798},
  primaryclass = {quant-ph},
  publisher = {arXiv},
  doi = {10.48550/arXiv.2503.14798},
  urldate = {2025-03-25},
  archiveprefix = {arXiv},
  langid = {english}
}

@article{calusine_analysis_2018,
  title = {Analysis and Mitigation of Interface Losses in Trenched Superconducting Coplanar Waveguide Resonators},
  author = {Calusine, G. and Melville, A. and Woods, W. and Das, R. and Stull, C. and Bolkhovsky, V. and Braje, D. and Hover, D. and Kim, D. K. and Miloshi, X. and Rosenberg, D. and Sevi, A. and Yoder, J. L. and Dauler, E. and Oliver, W. D.},
  year = 2018,
  month = feb,
  journal = {Applied Physics Letters},
  volume = {112},
  number = {6},
  pages = {062601},
  issn = {0003-6951, 1077-3118},
  doi = {10.1063/1.5006888},
  urldate = {2021-02-15},
  langid = {english}
}

@article{chang_eliminating_2025,
  title = {Eliminating {{Surface Oxides}} of {{Superconducting Circuits}} with {{Noble Metal Encapsulation}}},
  author = {Chang, Ray D. and Shumiya, Nana and McLellan, Russell A. and Zhang, Yifan and Bland, Matthew P. and Bahrami, Faranak and Mun, Junsik and Zhou, Chenyu and Kisslinger, Kim and Cheng, Guangming and Smitham, Basil M. and {Pakpour-Tabrizi}, Alexander C. and Yao, Nan and Zhu, Yimei and Liu, Mingzhao and Cava, Robert J. and Gopalakrishnan, Sarang and Houck, Andrew A. and De Leon, Nathalie P.},
  year = 2025,
  month = mar,
  journal = {Physical Review Letters},
  volume = {134},
  number = {9},
  pages = {097001},
  issn = {0031-9007, 1079-7114},
  doi = {10.1103/PhysRevLett.134.097001},
  urldate = {2025-04-28},
  langid = {english}
}

@article{chernyshov_frequency_2016,
  title = {Frequency Analysis for Modulation-Enhanced Powder Diffraction},
  author = {Chernyshov, Dmitry and Dyadkin, Vadim and Van Beek, Wouter and Urakawa, Atsushi},
  year = 2016,
  month = jul,
  journal = {Acta Crystallographica Section A Foundations and Advances},
  volume = {72},
  number = {4},
  pages = {500--506},
  issn = {2053-2733},
  doi = {10.1107/S2053273316008378},
  urldate = {2025-02-04}
}

@misc{crowley_disentangling_2023,
  title = {Disentangling {{Losses}} in {{Tantalum Superconducting Circuits}}},
  author = {Crowley, Kevin D. and McLellan, Russell A. and Dutta, Aveek and Shumiya, Nana and Place, Alexander P. M. and Le, Xuan Hoang and Gang, Youqi and Madhavan, Trisha and Khedkar, Nishaad and Feng, Yiming Cady and Umbarkar, Esha A. and Gui, Xin and Rodgers, Lila V. H. and Jia, Yichen and Feldman, Mayer M. and Lyon, Stephen A. and Liu, Mingzhao and Cava, Robert J. and Houck, Andrew A. and {de Leon}, Nathalie P.},
  year = 2023,
  month = jan,
  number = {arXiv:2301.07848},
  eprint = {2301.07848},
  primaryclass = {cond-mat, physics:quant-ph},
  publisher = {arXiv},
  urldate = {2023-01-24},
  archiveprefix = {arXiv}
}

@article{de_graaf_suppression_2018,
  title = {Suppression of Low-Frequency Charge Noise in Superconducting Resonators by Surface Spin Desorption},
  author = {{de Graaf}, S. E. and Faoro, L. and Burnett, J. and Adamyan, A. A. and Tzalenchuk, A. {\relax Ya}. and Kubatkin, S. E. and Lindstr{\"o}m, T. and Danilov, A. V.},
  year = 2018,
  month = mar,
  journal = {Nature Communications},
  volume = {9},
  number = {1},
  pages = {1143},
  issn = {2041-1723},
  doi = {10.1038/s41467-018-03577-2}
}

@article{face_nucleation_1987,
  title = {Nucleation of Body-Centered-Cubic Tantalum Films with a Thin Niobium Underlayer},
  author = {Face, D. W. and Prober, D. E.},
  year = 1987,
  month = nov,
  journal = {Journal of Vacuum Science \& Technology A: Vacuum, Surfaces, and Films},
  volume = {5},
  number = {6},
  pages = {3408--3411},
  issn = {0734-2101, 1520-8559},
  doi = {10.1116/1.574203},
  urldate = {2025-01-24},
  langid = {english}
}

@article{fairley_systematic_2021,
  title = {Systematic and Collaborative Approach to Problem Solving Using {{X-ray}} Photoelectron Spectroscopy},
  author = {Fairley, Neal and Fernandez, Vincent and Richard-Plouet, Mireille and {Guillot-Deudon}, Catherine and Walton, John and Smith, Emily and Flahaut, Delphine and Greiner, Mark and Biesinger, Mark and Tougaard, Sven and Morgan, David and Baltrusaitis, Jonas},
  year = 2021,
  month = sep,
  journal = {Applied Surface Science Advances},
  volume = {5},
  pages = {100112},
  issn = {2666-5239},
  doi = {10.1016/j.apsadv.2021.100112},
  urldate = {2022-05-31},
  langid = {english}
}

@article{ganjam_surpassing_2024,
  title = {Surpassing Millisecond Coherence in on Chip Superconducting Quantum Memories by Optimizing Materials and Circuit Design},
  author = {Ganjam, Suhas and Wang, Yanhao and Lu, Yao and Banerjee, Archan and Lei, Chan U and Krayzman, Lev and Kisslinger, Kim and Zhou, Chenyu and Li, Ruoshui and Jia, Yichen and Liu, Mingzhao and Frunzio, Luigi and Schoelkopf, Robert J.},
  year = 2024,
  month = may,
  journal = {Nature Communications},
  volume = {15},
  number = {1},
  pages = {3687},
  issn = {2041-1723},
  doi = {10.1038/s41467-024-47857-6},
  urldate = {2024-12-06},
  langid = {english}
}

@phdthesis{gao_physics_2008,
  title = {The {{Physics}} of {{Superconducting Microwave Resonators}}},
  author = {Gao, Jiansong},
  year = 2008,
  langid = {english},
  school = {California Institute of Technology,}
}

@misc{goronzy_comparison_2025,
  title = {Comparison of {{Nb}} and {{Ta Pentoxide Loss Tangents}} for {{Superconducting Quantum Devices}}},
  author = {Goronzy, D. P. and Mah, W. W. and Lim, P. G. and Guess, T. and Majumder, S. and {Garcia-Wetten}, D. A. and Walker, M. J. and Ramirez, J. and Syong, W.-R. and Bennett, D. and Vissers, M. and dos Reis, R. and Pham, T. and Dravid, V. P. and Hersam, M. C. and Bedzyk, M. J. and McRae, C. R. H.},
  year = 2025,
  month = dec,
  number = {arXiv:2512.05407},
  eprint = {2512.05407},
  primaryclass = {quant-ph},
  publisher = {arXiv},
  doi = {10.48550/arXiv.2512.05407},
  urldate = {2025-12-16},
  archiveprefix = {arXiv}
}

@article{guo_near-field_2023,
  title = {Near-{{Field Localization}} of the {{Boson Peak}} on {{Tantalum Films}} for {{Superconducting Quantum Devices}}},
  author = {Guo, Xiao and Degnan, Zachary and Steele, Julian A. and Solano, Eduardo and Donose, Bogdan C. and Bertling, Karl and Fedorov, Arkady and Raki{\'c}, Aleksandar D. and Jacobson, Peter},
  year = 2023,
  month = may,
  journal = {The Journal of Physical Chemistry Letters},
  volume = {14},
  number = {20},
  pages = {4892--4900},
  issn = {1948-7185, 1948-7185},
  doi = {10.1021/acs.jpclett.3c00850},
  urldate = {2025-01-30},
  copyright = {https://doi.org/10.15223/policy-029},
  langid = {english}
}

@article{jacob_update_2009,
  title = {An Update on the Thermodynamics of {{Ta2O5}}},
  author = {Jacob, K.T. and Shekhar, Chander and Waseda, Y.},
  year = 2009,
  month = jun,
  journal = {The Journal of Chemical Thermodynamics},
  volume = {41},
  number = {6},
  pages = {748--753},
  publisher = {Elsevier BV},
  issn = {0021-9614},
  doi = {10.1016/j.jct.2008.12.006},
  urldate = {2025-07-24},
  copyright = {https://www.elsevier.com/tdm/userlicense/1.0/},
  langid = {english}
}

@article{joiner_cleaning_2014,
  title = {Cleaning Graphene with a Titanium Sacrificial Layer},
  author = {Joiner, C. A. and Roy, T. and Hesabi, Z. R. and Chakrabarti, B. and Vogel, E. M.},
  year = 2014,
  month = jun,
  journal = {Applied Physics Letters},
  volume = {104},
  number = {22},
  publisher = {AIP Publishing},
  issn = {0003-6951, 1077-3118},
  doi = {10.1063/1.4881886},
  urldate = {2025-07-23},
  langid = {english}
}

@article{kalboussi_reducing_2024,
  title = {Reducing Two-Level Systems Dissipations in {{3D}} Superconducting Niobium Resonators by Atomic Layer Deposition and High Temperature Heat Treatment},
  author = {Kalboussi, Y. and Delatte, B. and Bira, S. and Dembele, K. and Li, X. and Miserque, F. and Brun, N. and Walls, M. and Maurice, J. L. and Dragoe, D. and Leroy, J. and Longuevergne, D. and Gentils, A. and {Jublot-Leclerc}, S. and Jullien, G. and Eozenou, F. and Baudrier, M. and Maurice, L. and Proslier, T.},
  year = 2024,
  month = mar,
  journal = {Applied Physics Letters},
  volume = {124},
  number = {13},
  pages = {134001},
  issn = {0003-6951, 1077-3118},
  doi = {10.1063/5.0202214},
  urldate = {2025-04-14},
  langid = {english}
}

@article{lozano_low-loss_2024,
  title = {Low-Loss Alpha-Tantalum Coplanar Waveguide Resonators on Silicon Wafers: Fabrication, Characterization and Surface Modification},
  shorttitle = {Low-Loss Alpha-Tantalum Coplanar Waveguide Resonators on Silicon Wafers},
  author = {Lozano, Daniel P{\'e}rez and Mongillo, Massimo and Piao, Xiaoyu and Couet, Sebastien and Wan, Danny and Canvel, Yann and Vadiraj, A. M. and Ivanov, Tsvetan and Verjauw, Jeroen and Acharya, Rohith and Van Damme, Jacques and Fahd, Mohiyaddin A. and Jussot, Julien and Gowda, Pallavi Puttarame and Pacco, Antoine and Raes, Bart and Van De Vondel, Joris and Radu, Iuliana and Govoreanu, Bogdan and Swerts, Johan and Potocnik, Anton and DeGreve, Kristiaan},
  year = 2024,
  month = may,
  journal = {Materials for Quantum Technology},
  issn = {2633-4356},
  doi = {10.1088/2633-4356/ad4b8c},
  urldate = {2024-05-17},
  copyright = {https://creativecommons.org/licenses/by/4.0/},
  langid = {english}
}

@misc{lozano_manufacturing_2022,
  title = {Manufacturing High-{{Q}} Superconducting Alpha-Tantalum Resonators on Silicon Wafers},
  author = {Lozano, D. P. and Mongillo, M. and Piao, X. and Couet, S. and Wan, D. and Canvel, Y. and Vadiraj, A. M. and Ivanov, Ts and Verjauw, J. and Acharya, R. and Van Damme, J. and Mohiyaddin, F. A. and Jussot, J. and Gowda, P. P. and Pacco, A. and Raes, B. and {Van de Vondel}, J. and Radu, I. P. and Govoreanu, B. and Swerts, J. and Poto{\v c}nik, A. and De Greve, K.},
  year = 2022,
  month = nov,
  number = {arXiv:2211.16437},
  eprint = {2211.16437},
  primaryclass = {cond-mat, physics:physics, physics:quant-ph},
  publisher = {arXiv},
  urldate = {2022-12-06},
  archiveprefix = {arXiv}
}

@misc{marcaud_low-loss_2025,
  title = {Low-{{Loss Superconducting Resonators Fabricated}} from {{Tantalum Films Grown}} at {{Room Temperature}}},
  author = {Marcaud, Guillaume and Perello, David and Chen, Cliff and Umbarkar, Esha and Weiland, Conan and Gao, Jiansong and Diez, Sandra and Ly, Victor and Mahuli, Neha and D'Souza, Nathan and He, Yuan and Aghaeimeibodi, Shahriar and Resnick, Rachel and Jaye, Cherno and Rumaiz, Abdul K. and Fischer, Daniel A. and Hunt, Matthew and Painter, Oskar and Jarrige, Ignace},
  year = 2025,
  month = jan,
  number = {arXiv:2501.09885},
  eprint = {2501.09885},
  primaryclass = {physics},
  publisher = {arXiv},
  doi = {10.48550/arXiv.2501.09885},
  urldate = {2025-01-22},
  archiveprefix = {arXiv}
}

@article{martinis_surface_2022,
  title = {Surface Loss Calculations and Design of a Superconducting Transmon Qubit with Tapered Wiring},
  author = {Martinis, John M.},
  year = 2022,
  month = mar,
  journal = {npj Quantum Information},
  volume = {8},
  number = {1},
  pages = {26},
  issn = {2056-6387},
  doi = {10.1038/s41534-022-00530-6},
  urldate = {2024-05-24},
  langid = {english}
}

@misc{mclellan_chemical_2023,
  title = {Chemical Profiles of the Oxides on Tantalum in State of the Art Superconducting Circuits},
  author = {McLellan, Russell A. and Dutta, Aveek and Zhou, Chenyu and Jia, Yichen and Weiland, Conan and Gui, Xin and Place, Alexander P. M. and Crowley, Kevin D. and Le, Xuan Hoang and Madhavan, Trisha and Gang, Youqi and Baker, Lukas and Head, Ashley R. and Waluyo, Iradwikanari and Li, Ruoshui and Kisslinger, Kim and Hunt, Adrian and Jarrige, Ignace and Lyon, Stephen A. and Barbour, Andi M. and Cava, Robert J. and Houck, Andrew A. and Hulbert, Steven L. and Liu, Mingzhao and Walter, Andrew L. and {de Leon}, Nathalie P.},
  year = 2023,
  month = jan,
  number = {arXiv:2301.04567},
  eprint = {2301.04567},
  primaryclass = {cond-mat, physics:quant-ph},
  publisher = {arXiv},
  urldate = {2023-01-16},
  archiveprefix = {arXiv}
}

@article{mcrae_materials_2020,
  title = {Materials Loss Measurements Using Superconducting Microwave Resonators},
  author = {McRae, C. R. H. and Wang, H. and Gao, J. and Vissers, M. R. and Brecht, T. and Dunsworth, A. and Pappas, D. P. and Mutus, J.},
  year = 2020,
  month = sep,
  journal = {Review of Scientific Instruments},
  volume = {91},
  number = {9},
  pages = {091101},
  issn = {0034-6748, 1089-7623},
  doi = {10.1063/5.0017378},
  urldate = {2022-01-31},
  langid = {english}
}

@article{müller_understanding_2019,
  title = {Towards Understanding Two-Level-Systems in Amorphous Solids: Insights from Quantum Circuits},
  shorttitle = {Towards Understanding Two-Level-Systems in Amorphous Solids},
  author = {M{\"u}ller, Clemens and Cole, Jared H and Lisenfeld, J{\"u}rgen},
  year = 2019,
  month = dec,
  journal = {Reports on Progress in Physics},
  volume = {82},
  number = {12},
  pages = {124501},
  issn = {0034-4885, 1361-6633},
  doi = {10.1088/1361-6633/ab3a7e},
  urldate = {2020-11-24},
  langid = {english}
}

@article{murray_material_2021,
  title = {Material Matters in Superconducting Qubits},
  author = {Murray, Conal E.},
  year = 2021,
  month = oct,
  journal = {Materials Science and Engineering: R: Reports},
  volume = {146},
  pages = {100646},
  issn = {0927796X},
  doi = {10.1016/j.mser.2021.100646},
  urldate = {2022-01-19},
  langid = {english}
}

@article{place_new_2021,
  title = {New Material Platform for Superconducting Transmon Qubits with Coherence Times Exceeding 0.3 Milliseconds},
  author = {Place, Alexander P. M. and Rodgers, Lila V. H. and Mundada, Pranav and Smitham, Basil M. and Fitzpatrick, Mattias and Leng, Zhaoqi and Premkumar, Anjali and Bryon, Jacob and Vrajitoarea, Andrei and Sussman, Sara and Cheng, Guangming and Madhavan, Trisha and Babla, Harshvardhan K. and Le, Xuan Hoang and Gang, Youqi and J{\"a}ck, Berthold and Gyenis, Andr{\'a}s and Yao, Nan and Cava, Robert J. and {de Leon}, Nathalie P. and Houck, Andrew A.},
  year = 2021,
  month = dec,
  journal = {Nature Communications},
  volume = {12},
  number = {1},
  pages = {1779},
  issn = {2041-1723},
  doi = {10.1038/s41467-021-22030-5},
  urldate = {2022-05-03},
  langid = {english}
}

@article{pritchard_suppressed_2025,
  title = {Suppressed Paramagnetism in Amorphous {{Ta}} 2 {{O}} 5 - x Oxides and Its Link to Superconducting-Qubit Performance},
  author = {Pritchard, P. Graham and Rondinelli, James M.},
  year = 2025,
  month = jun,
  journal = {Physical Review Applied},
  volume = {23},
  number = {6},
  pages = {064062},
  issn = {2331-7019},
  doi = {10.1103/6wyt-fxjg},
  urldate = {2025-12-16},
  langid = {english}
}

@article{probst_efficient_2015,
  title = {Efficient and Robust Analysis of Complex Scattering Data under Noise in Microwave Resonators},
  author = {Probst, S. and Song, F. B. and Bushev, P. A. and Ustinov, A. V. and Weides, M.},
  year = 2015,
  month = feb,
  journal = {Review of Scientific Instruments},
  volume = {86},
  number = {2},
  pages = {024706},
  issn = {0034-6748, 1089-7623},
  doi = {10.1063/1.4907935},
  urldate = {2020-11-24},
  langid = {english}
}

@article{ren_annealing_2021,
  title = {Annealing Effects on the Optical and Electrochemical Properties of Tantalum Pentoxide Films},
  author = {Ren, Wei and Yang, Guang-Dao and Feng, Ai-Ling and Miao, Rui-Xia and Xia, Jun-Bo and Wang, Yong-Gang},
  year = 2021,
  month = aug,
  journal = {Journal of Advanced Ceramics},
  volume = {10},
  number = {4},
  pages = {704--713},
  issn = {2226-4108, 2227-8508},
  doi = {10.1007/s40145-021-0465-2},
  urldate = {2024-10-16},
  langid = {english}
}

@article{rieger_fano_2023,
  title = {Fano {{Interference}} in {{Microwave Resonator Measurements}}},
  author = {Rieger, D. and G{\"u}nzler, S. and Spiecker, M. and Nambisan, A. and Wernsdorfer, W. and Pop, I. M.},
  year = 2023,
  month = jul,
  journal = {Physical Review Applied},
  volume = {20},
  number = {1},
  eprint = {2209.03036},
  primaryclass = {quant-ph},
  pages = {014059},
  issn = {2331-7019},
  doi = {10.1103/PhysRevApplied.20.014059},
  urldate = {2023-08-23},
  archiveprefix = {arXiv},
  langid = {english}
}

@article{shi_tantalum_2022,
  title = {Tantalum Microwave Resonators with Ultra-High Intrinsic Quality Factors},
  author = {Shi, Lili and Guo, Tingting and Su, Runfeng and Chi, Tianyuan and Sheng, Yifan and Jiang, Junliang and Cao, Chunhai and Wu, Jingbo and Tu, Xuecou and Sun, Guozhu and Chen, Jian and Wu, Peiheng},
  year = 2022,
  month = dec,
  journal = {Applied Physics Letters},
  volume = {121},
  number = {24},
  pages = {242601},
  issn = {0003-6951, 1077-3118},
  doi = {10.1063/5.0124821},
  urldate = {2022-12-15},
  langid = {english}
}

@misc{sommers_open-source_2025,
  title = {Open-{{Source Highly Parallel Electromagnetic Simulations}} for {{Superconducting Circuits}}},
  author = {Sommers, David and Degnan, Zach and Gautam, Divita and Chen, Yi-Hsun and Chiu, Chun-Ching and Fedorov, Arkady and Pakkiam, Prasanna},
  year = 2025,
  publisher = {arXiv},
  doi = {10.48550/ARXIV.2511.01220},
  urldate = {2025-12-01},
  copyright = {Creative Commons Attribution 4.0 International}
}

@article{tuokkola_methods_2025,
  title = {Methods to Achieve Near-Millisecond Energy Relaxation and Dephasing Times for a Superconducting Transmon Qubit},
  author = {Tuokkola, Mikko and Sunada, Yoshiki and Kivij{\"a}rvi, Heidi and Albanese, Jonatan and Gr{\"o}nberg, Leif and Kaikkonen, Jukka-Pekka and Vesterinen, Visa and Govenius, Joonas and M{\"o}tt{\"o}nen, Mikko},
  year = 2025,
  month = feb,
  eprint = {2407.18778},
  primaryclass = {quant-ph},
  doi = {10.48550/arXiv.2407.18778},
  urldate = {2025-02-24},
  archiveprefix = {arXiv},
  langid = {english}
}

@article{urade_microwave_2024,
  title = {Microwave Characterization of Tantalum Superconducting Resonators on Silicon Substrate with Niobium Buffer Layer},
  author = {Urade, Yoshiro and Yakushiji, Kay and Tsujimoto, Manabu and Yamada, Takahiro and Makise, Kazumasa and Mizubayashi, Wataru and Inomata, Kunihiro},
  year = 2024,
  month = feb,
  journal = {APL Materials},
  volume = {12},
  number = {2},
  pages = {021132},
  issn = {2166-532X},
  doi = {10.1063/5.0165137},
  urldate = {2024-10-11},
  langid = {english}
}

@article{van_damme_argon-milling-induced_2023,
  title = {Argon-{{Milling-Induced Decoherence Mechanisms}} in {{Superconducting Quantum Circuits}}},
  author = {Van Damme, J. and Ivanov, {\relax Ts}. and Favia, P. and Conard, T. and Verjauw, J. and Acharya, R. and Perez Lozano, D. and Raes, B. and Van De Vondel, J. and Vadiraj, A.M. and Mongillo, M. and Wan, D. and De Boeck, J. and Poto{\v c}nik, A. and De Greve, K.},
  year = 2023,
  month = jul,
  journal = {Physical Review Applied},
  volume = {20},
  number = {1},
  pages = {014034},
  issn = {2331-7019},
  doi = {10.1103/PhysRevApplied.20.014034},
  urldate = {2025-07-31},
  langid = {english}
}

@article{verjauw_investigation_2021,
  title = {Investigation of {{Microwave Loss Induced}} by {{Oxide Regrowth}} in {{High-}} {{{\emph{Q}}}} {{Niobium Resonators}}},
  author = {Verjauw, J. and Poto{\v c}nik, A. and Mongillo, M. and Acharya, R. and Mohiyaddin, F. and Simion, G. and Pacco, A. and Ivanov, {\relax Ts}. and Wan, D. and Vanleenhove, A. and Souriau, L. and Jussot, J. and Thiam, A. and Swerts, J. and Piao, X. and Couet, S. and Heyns, M. and Govoreanu, B. and Radu, I.},
  year = 2021,
  month = jul,
  journal = {Physical Review Applied},
  volume = {16},
  number = {1},
  pages = {014018},
  issn = {2331-7019},
  doi = {10.1103/PhysRevApplied.16.014018},
  urldate = {2022-12-16},
  langid = {english}
}

@article{wang_practical_2022,
  title = {Towards Practical Quantum Computers: Transmon Qubit with a Lifetime Approaching 0.5 Milliseconds},
  shorttitle = {Towards Practical Quantum Computers},
  author = {Wang, Chenlu and Li, Xuegang and Xu, Huikai and Li, Zhiyuan and Wang, Junhua and Yang, Zhen and Mi, Zhenyu and Liang, Xuehui and Su, Tang and Yang, Chuhong and Wang, Guangyue and Wang, Wenyan and Li, Yongchao and Chen, Mo and Li, Chengyao and Linghu, Kehuan and Han, Jiaxiu and Zhang, Yingshan and Feng, Yulong and Song, Yu and Ma, Teng and Zhang, Jingning and Wang, Ruixia and Zhao, Peng and Liu, Weiyang and Xue, Guangming and Jin, Yirong and Yu, Haifeng},
  year = 2022,
  month = dec,
  journal = {npj Quantum Information},
  volume = {8},
  number = {1},
  pages = {3},
  issn = {2056-6387},
  doi = {10.1038/s41534-021-00510-2},
  urldate = {2022-02-09},
  langid = {english}
}

@article{wang_reducing_2017,
  title = {Reducing Graphene Device Variability with Yttrium Sacrificial Layers},
  author = {Wang, Ning C. and Carrion, Enrique A. and Tung, Maryann C. and Pop, Eric},
  year = 2017,
  month = may,
  journal = {Applied Physics Letters},
  volume = {110},
  number = {22},
  publisher = {AIP Publishing},
  issn = {0003-6951, 1077-3118},
  doi = {10.1063/1.4984090},
  urldate = {2025-07-10},
  langid = {english}
}

@article{wang_surface_2015,
  ids = {wang2015a},
  title = {Surface Participation and Dielectric Loss in Superconducting Qubits},
  author = {Wang, C. and Axline, C. and Gao, Y. Y. and Brecht, T. and Chu, Y. and Frunzio, L. and Devoret, M. H. and Schoelkopf, R. J.},
  year = 2015,
  month = oct,
  journal = {Applied Physics Letters},
  volume = {107},
  number = {16},
  pages = {162601},
  issn = {0003-6951, 1077-3118},
  doi = {10.1063/1.4934486},
  urldate = {2020-08-24},
  langid = {english}
}

@article{woods_determining_2019,
  title = {Determining {{Interface Dielectric Losses}} in {{Superconducting Coplanar-Waveguide Resonators}}},
  author = {Woods, W. and Calusine, G. and Melville, A. and Sevi, A. and Golden, E. and Kim, D.K. and Rosenberg, D. and Yoder, J.L. and Oliver, W.D.},
  year = 2019,
  month = jul,
  journal = {Physical Review Applied},
  volume = {12},
  number = {1},
  pages = {014012},
  issn = {2331-7019},
  doi = {10.1103/PhysRevApplied.12.014012},
  urldate = {2021-02-15},
  langid = {english}
}

@article{xu_high_2008,
  title = {High Temperature Annealing Effect on Structure, Optical Property and Laser-Induced Damage Threshold of {{Ta2O5}} Films},
  author = {Xu, Cheng and Xiao, Qiling and Ma, Jianyong and Jin, Yunxia and Shao, Jianda and Fan, Zhengxiu},
  year = 2008,
  month = aug,
  journal = {Applied Surface Science},
  volume = {254},
  number = {20},
  pages = {6554--6559},
  issn = {01694332},
  doi = {10.1016/j.apsusc.2008.04.034},
  urldate = {2024-10-16},
  copyright = {https://www.elsevier.com/tdm/userlicense/1.0/},
  langid = {english}
}

@article{zhou_ultrathin_2024,
  title = {Ultrathin {{Magnesium-Based Coating}} as an {{Efficient Oxygen Barrier}} for {{Superconducting Circuit Materials}}},
  author = {Zhou, Chenyu and Mun, Junsik and Yao, Juntao and kumar Anbalagan, Aswin and Hossain, Mohammad D. and McLellan, Russell A. and Li, Ruoshui and Kisslinger, Kim and Li, Gengnan and Tong, Xiao and Head, Ashley R. and Weiland, Conan and Hulbert, Steven L. and Walter, Andrew L. and Li, Qiang and Zhu, Yimei and Sushko, Peter V. and Liu, Mingzhao},
  year = 2024,
  journal = {Advanced Materials},
  volume = {36},
  number = {18},
  pages = {2310280},
  issn = {1521-4095},
  doi = {10.1002/adma.202310280},
  urldate = {2025-12-11},
  copyright = {\copyright{} 2024 Wiley-VCH GmbH},
  langid = {english}
}

\appendix
\clearpage
\newpage

\section{Fabrication methods}
To construct our Ti-Ta-Si material stack, we deposit a 2~Å sacrificial Ti layer atop a bcc $\alpha$-Ta film, which is known to host fewer TLS defects than the tetragonal $\beta$-Ta phase.~\cite{place_new_2021}
A 200~nm $\alpha$-Ta film is grown by magnetron sputtering on high-resistivity Si at room temperature using a Nb buffer layer, which promotes high-quality single-phase $\alpha$-Ta growth due to its minimal ($<0.1\%$) lattice mismatch with Ta.~\cite{urade_microwave_2024, face_nucleation_1987, marcaud_low-loss_2025}
The film phase is verified by synchrotron grazing-incidence wide-angle X-ray scattering (GIWAXS) measurements (Figs.~\ref{fig:giwaxs},~\ref{fig:giwaxs_peaks}), where strong Bragg peaks corresponding to \textit{bcc} $\alpha$-Ta are observed and no peaks from $\beta$-Ta are detected.
The sheet resistance of the $\alpha$-Ta film was measured using a Hall bar geometry to be $R_\square=2.2 \ \Omega/\square$, corresponding to a kinetic inductance of $L_K=0.64$~pH$/\square$.

To prepare the wafers for fabrication, 8" Ta-on-Si substrates are coated with 2~µm of AZ1512~HS photoresist for protection before dicing into $20\times20$~mm$^2$ chiplets.
After dicing, the photoresist is removed by sequential 5~min sonications in acetone and isopropyl alcohol (IPA).
Bare Ta chips are then loaded into the load lock chamber of an electron beam evaporator (PLASSYS~MEB-550S), which is pumped to a base pressure of $4.8\times10^{-8}$~mbar for \textit{in situ} ion milling, Ti deposition, and oxidation.

The native Ta oxide is first removed by 5.5~min of gentle ion beam milling using a Kaufman ion source with low-energy (400~eV) Ar\textsuperscript{+} ions at a flux density of 15~mA~cm\textsuperscript{-2}.
Without breaking vacuum, an ultrathin ($2\pm0.5$~Å) Ti layer is deposited by electron beam evaporation at 0.2~Å~s\textsuperscript{-1}, followed by \textit{in situ} oxidation for 10~min in 10~Torr of pure O\textsubscript{2}.
This Ti-coated sample is hereafter referred to as Ti/Ta.
We emphasise that this method does not require \textit{in situ} Ta and Ti co-deposition and is compatible with pre-existing Ta films.

For optical lithography, AZ1512~HS photoresist is spin-coated at 6000~RPM for 60~s (yielding 1.4~µm thickness), soft-baked at 110~$^\circ$C for 60~s, and allowed to cool to room temperature.
Exposure is performed with a Heidelberg MLA~150 maskless aligner (405~nm laser), followed by a post-exposure bake (110~$^\circ$C for 60~s).
The pattern is developed in AZ726~MIF for 30~s, then rinsed in two successive DI water baths (30~s each).
The developed resist serves as the etch mask for subsequent pattern transfer.
All samples contain five transmission-line coupled $\lambda/4$ resonators with a CPW width and gap of 9~µm and 5~µm respectively.

Etching is carried out in an Oxford Instruments PlasmaPro 80 RIE system using CF\textsubscript{4} plasma.
Prior to processing, a 20~min chamber pre-clean (50~sccm O\textsubscript{2}, 25~sccm CF\textsubscript{4}, 100~mTorr, 150~W) is performed.
The sample is then transferred into the chamber, which is pumped to $\sim5\times10^{-6}$~mbar.

A 10~s O\textsubscript{2} plasma pre-clean (40~sccm O\textsubscript{2}, 100~mTorr, 150~W) removes residual organics.
The Ti/Ta film is etched in CF\textsubscript{4} plasma for 20~min at a rate of $\sim11$~nm~min\textsuperscript{-1} (30~sccm CF\textsubscript{4}, 20~mTorr, 200~W).
Under these conditions, the underlying Si etches at $\sim22$~nm~min\textsuperscript{-1}, twice the rate of Ta.
Thus, the 200~nm Ta layer is expected to be fully removed in 18~min, with $\sim50$~nm of Si trenched in the remaining 2~min.
Given that the Ta etch rate may vary by up to $\pm1$~nm~min\textsuperscript{-1}, this margin ensures complete removal across all samples.
After etching, a 2~min post-clean in O\textsubscript{2} plasma (same parameters) helps remove fluorocarbon etch byproducts, which are known to contribute to TLS loss.~\cite{guo_near-field_2023, tuokkola_methods_2025}
Etch completion is verified optically, after which the sample is coated again with AZ1512~HS and diced into individual $2\times7$~mm\textsuperscript{2} circuits.

To remove the protective photoresist, a 5~min sonication in acetone and IPA is performed.
Residual organics and adsorbed fluorocarbons are removed by 5~min O\textsubscript{2} plasma ashing at 200~W RF power in a Diener Atto low-pressure plasma system.
Although this step is effective at residue removal, it leaves both Si and Ti/Ta surfaces heavily oxidised (see Supplementary Information).
Samples are then vacuum-sealed and transported for post-processing.

Next, a buffered oxide etch (BOE) is used to remove surface oxides and dissolve the 2~Å Ti layer.
Samples are transferred directly into a 6:1 BOE solution of NH\textsubscript{4}F:HF for 20~min.
After etching, the samples are rinsed in DI water (2$\times$~30~s) and sonicated in IPA for 3~min.

Samples are then transferred into the PLASSYS~MEB-550S chamber, mounted on a molybdenum (Mo) stage, and pumped overnight to $9\times10^{-8}$~mbar.
They are annealed at $700^\circ$C for 10~hr, with a ramp rate of $\sim9^\circ$C~min\textsuperscript{-1}, and allowed to cool naturally to $30^\circ$C over approximately one day.
Following annealing, samples undergo a final 1~min BOE step and rinse (DI water + IPA sonication).

Wire bonding and packaging are completed within 3~hr of the final BOE, minimising air exposure and suppressing regrowth of surface oxides before cryostat loading.

\begin{figure}[!h]
    \centering
    \includegraphics[width=0.9\linewidth]{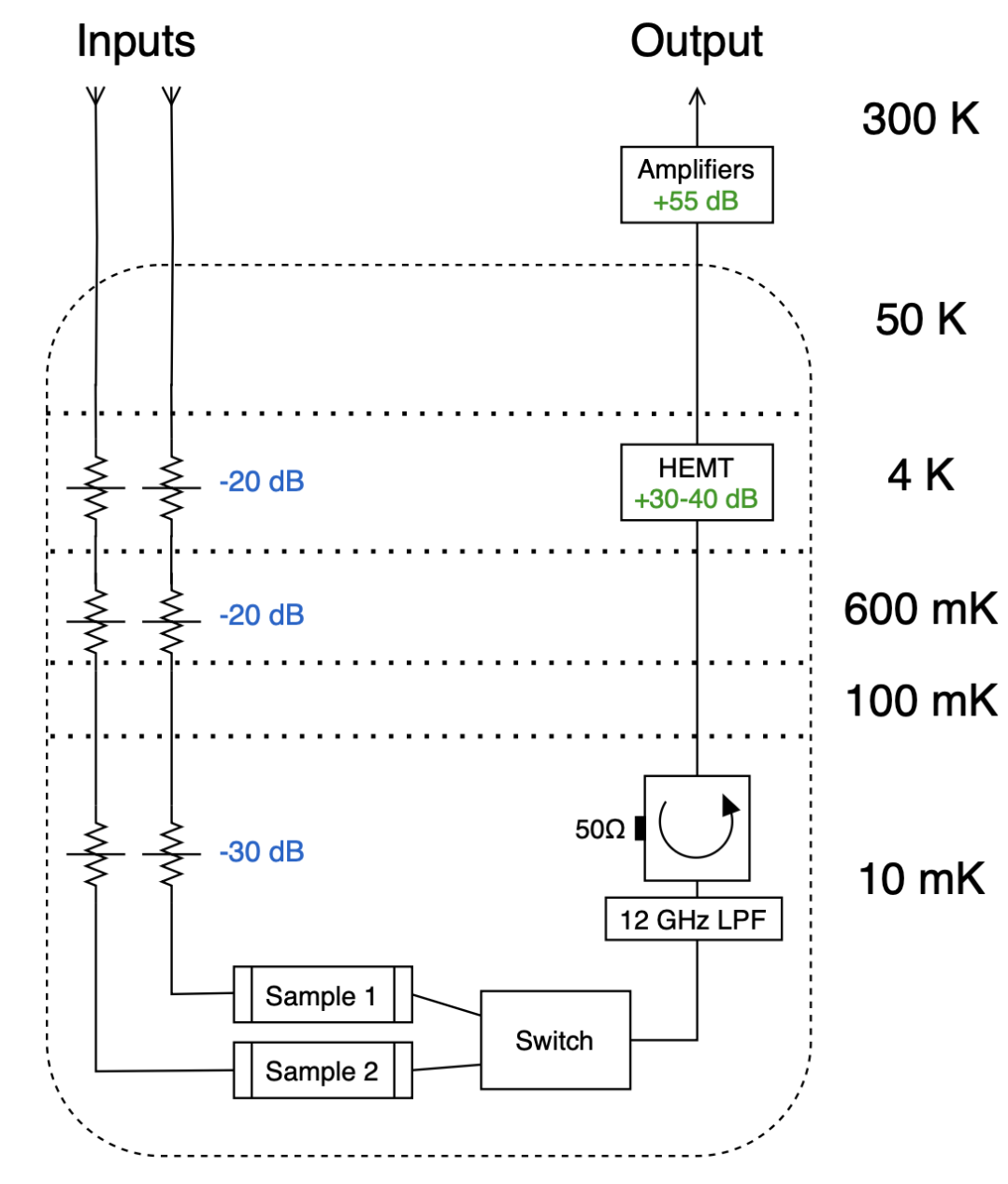}
    \caption{Wiring diagram for two-sample CPW resonator measurements.}
    \label{fig:fridge_wiring}
\end{figure}

\section{Resonator measurements}
For device measurement, the wire-bonded and packaged samples are shielded with mu-metal and aluminium cans, and mounted to the mixing chamber flange of a BlueFors SD dilution refrigerator, which reaches a base temperature of $\sim25$~mK. 
Two chips with five $\lambda/4$ resonators each are wire-bonded to a dual-slot PCB, allowing for the measurement of both in a single cooldown.
Separate input lines are used for each sample, and a cryogenic microwave switch (Radiall R5927B2141) is used to route output signals. A Keysight P9375A two-port vector network analyser (VNA) was used to measure the complex $S_{21}$ scattering signals.

\begin{figure*}[!h]
    \centering
    \includegraphics[width=0.8\linewidth]{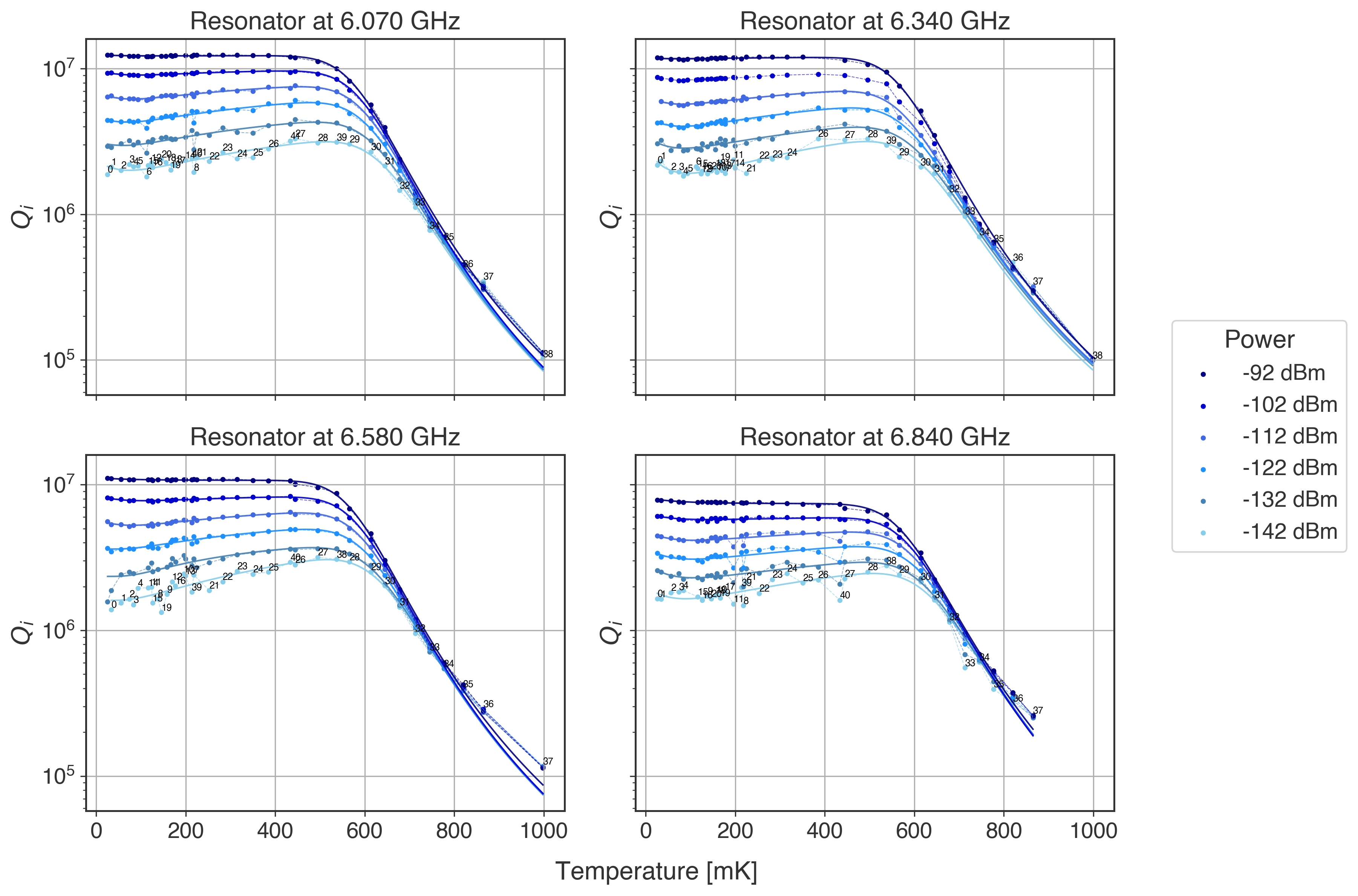}
    \caption{Temperature dependent $Q_i$ for four resonators on sample \texttt{T-LA2}. Experimental data is displayed as points with a fitted temperature dependent loss model displayed as solid lines. The power values shown in the legend are the applied powers at the on-chip transmission line, corresponding to a range of average intracavity photon numbers from $-142$~dBm~$\approx10^1$ photons to $-92$~dBm~$\approx10^6$ photons.}
    \label{fig:Qi_temp_dependence}
\end{figure*}

\subsubsection{Temperature dependent measurements}
The sample temperature was controlled via a 100~$\Omega$ resistive heater mounted to the same mixing chamber flange as the sample enclosure. %
A heating power of 4.9~mW was required to raise the flange temperature to approximately 1~K.
After each adjustment to the heater current, the flange temperature was given time to stabilise (typically within 1~hour), followed by an additional 1~hour to ensure thermal equilibrium of the sample. After thermalisation, the measurements corresponding to data-points in Fig.~\ref{fig:Qi_temp_dependence} were performed.

\section{XPS}
\begin{figure*}[!h]
    \centering
    \includegraphics[width=0.8\linewidth]{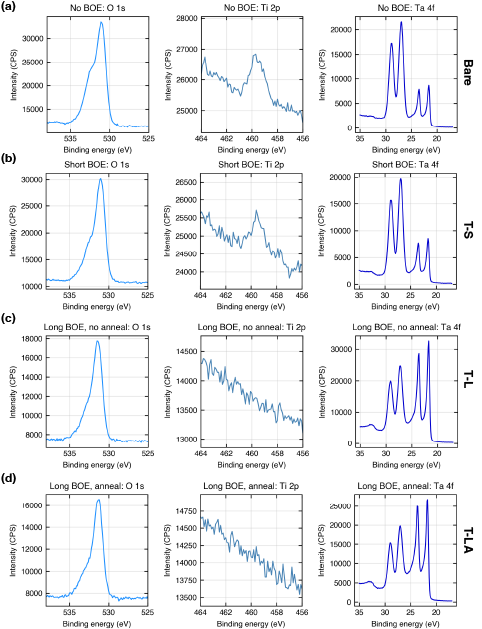}
    \caption{High resolution XPS scans of O~1s, Ti~2p, and Ta~4f levels tracking chemical changes for samples treated by: \textbf{(a)} no BOE (bare), \textbf{(b)} 1~min short BOE  without annealing (\texttt{T-S}), \textbf{(c)} 20~min long BOE without annealing (\texttt{T-L}), and \textbf{(d)} long BOE with annealing (\texttt{T-LA}). The data shows that the Ti layer remains in place after short BOE, and consists primarily of TiO$_2$. The Ti layer is removed by long BOE. The broad shoulder in the Ti~2p data around 468~eV is the Ta~4p$_{1/2}$. The XPS measurements were taken on samples that underwent identical fabrication processes to those with measured devices in the main text.}
    \label{fig:xps_all}
\end{figure*}

X-ray photoelectron spectroscopy (XPS) measurements were performed on a Kratos AXIS Supra+ system equipped with monochromated Al~K$\alpha$ ($1486.6$~eV) with beam spot size of $300$~$\mu$m. XPS chamber base pressure $<1\times 10^{-8}$ mbar. Survey scans ($0$ - $1200$~eV) and high-resolution scans of the Ta~4f, O~1s, C~1s, and Ti~2p levels were obtained (Fig.~\ref{fig:xps_all}). Samples were grounded by clips and no sample charging was observed. Component analysis was performed in CasaXPS, where peaks were fitted using Gaussian/Lorentzian line-shapes and Shirley background subtraction was used.~\cite{fairley_systematic_2021} The Ta~4f peak was fitted with doublet components from metallic Ta and Ta$_2$O$_5$. To use a minimal fitting model that accounts for details seen in angle resolved XPS data across all treatments, we implement an additional doublet at intermediate binding energy to account for suboxide contributions.

\section{GIWAXS}
\begin{figure*}[!h]
    \centering
    \includegraphics[width=0.9\linewidth]{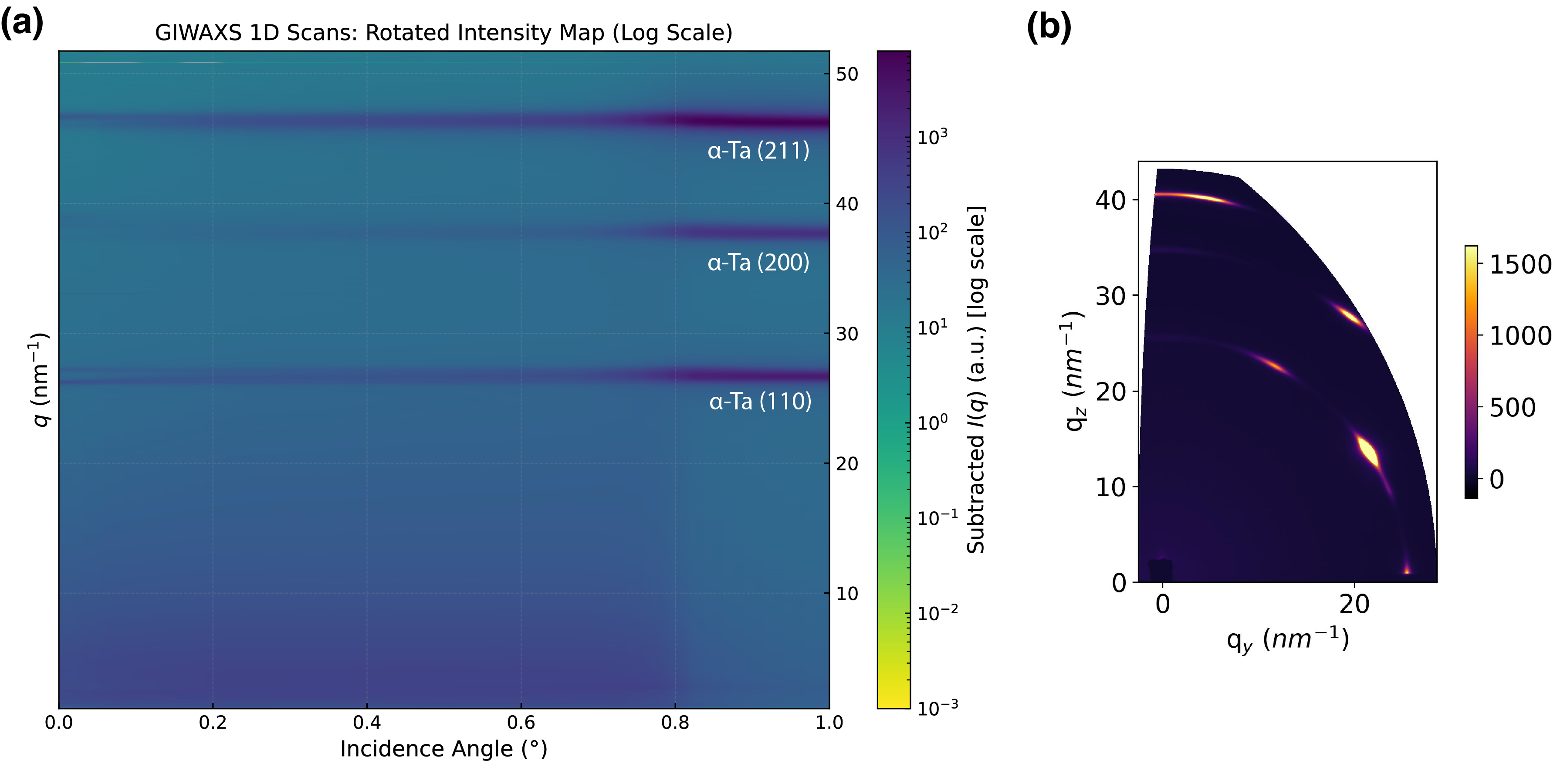}
    \caption{\textbf{(a)} 2D colour map of background-subtracted GIWAXS 1D scattering profiles for Ta resonators measured at varying incidence angles. The horizontal axis corresponds to the incidence angle (°), and the vertical axis shows the momentum transfer $q$ (nm\textsuperscript{-1}). Intensity is plotted on a logarithmic scale, with darker regions indicating higher scattered intensity. Air background was subtracted using a separately measured reference scan. \textbf{(b)} GIWAXS 2D detector image of the bulk Ta film at incident angle $\theta_i=0.90^\circ$ (above the critical angle, $\theta_c=0.27^\circ$. In-plane (IP) scattering is shown on the x-axis ($q_y$), and out-of-plane (OOP) on the y-axis ($q_z$).}
    \label{fig:giwaxs}
\end{figure*}

\begin{figure*}
    \centering
    \includegraphics[width=0.9\linewidth]{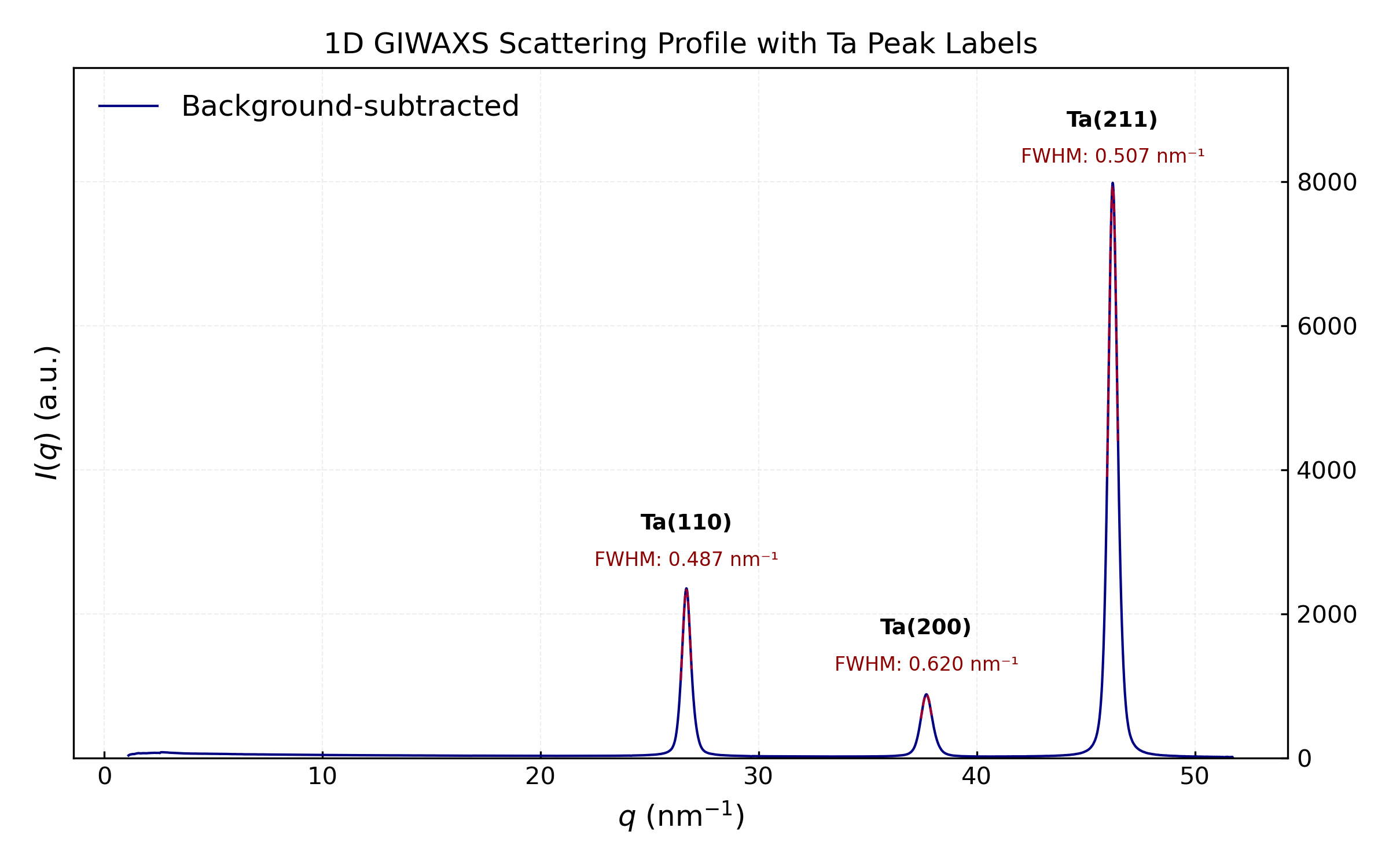}
    \caption{Bragg peaks with labelled full width half maximum (FWHM) from a background-subtracted GIWAXS scan of a bare Ta film at an incident angle of $\theta_i=0.9^\circ$.}
    \label{fig:giwaxs_peaks}
\end{figure*}

Grazing incidence wide angle X-ray scattering (GIWAXS) data, as shown in Figures \ref{fig:giwaxs} and \ref{fig:giwaxs_peaks}, was taken at the \href{https://www.cells.es/en/instruments/beamlines/bl11-ncd-sweet}{NCD-SWEET beamline (ALBA Synchrotron, Barcelona)} with an incident X-ray energy of 9.25~keV ($\lambda=1.34$~Å) and a beam spotsize FWHM $(x,y)= (150$~µm$,25$~µm). GIWAXS data was visualised and analysed using \href{https://soft.snbl.eu/medved/medved.html}{Medved software}.~\cite{chernyshov_frequency_2016}

\end{document}